\documentclass[prl,twocolumn,superscriptaddress,showpacs,amscd,amsmath,amssymb,verbatim]{revtex4}

\usepackage[dvipsnames]{xcolor}
\usepackage{graphicx}
\usepackage{textcomp}
\usepackage[OT1,T1]{fontenc}
\usepackage[squaren, Gray, cdot]{SIunits}
\usepackage{amsmath}
\usepackage{nicefrac}
\usepackage{natbib}
\usepackage{revsymb}
\usepackage[colorlinks=true,linkcolor=blue,citecolor=blue,urlcolor=blue]{hyperref}
\usepackage{ulem}
\usepackage{threeparttable}
\usepackage{multirow}
\usepackage{dcolumn}
\usepackage{booktabs}
\usepackage{tabularx}
\usepackage{caption} 

\newcommand{\SNIO}{Sr$_3$NiIrO$_6$}
\newcommand{\co}{(Color online) }
\newcommand{\K}{~$\kelvin$}
\newcommand{\T}{~$\tesla$}
\newcommand{\etal}{{\it et al.}}
\newcommand{\eV}{~$\electronvolt$}
\newcommand{\keV}{~$\kilo\electronvolt$}
\newcommand{\meV}{~$\milli\electronvolt$}

\newcommand{\tg}{$t_{2g}$}
\newcommand{\eg}{$e_g$}
\newcommand{\Tg}{$^2T_{2g}$}

\begin{document}

\title{Anisotropic interactions opposing magnetocrystalline anisotropy  in \SNIO}

\author{E. Lefran\c{c}ois}
\email[Corresponding author: ]{lefrancois@ill.fr}
\affiliation{Institut Laue Langevin,  CS 20156, 38042 Grenoble Cedex 9, France}
\affiliation{CNRS, Institut N\'eel, 38042 Grenoble, France}
\affiliation{Univ. Grenoble Alpes, Institut N\'eel, 38042 Grenoble, France}
\author{A.-M. Pradipto}
\email[Corresponding author: ]{a.m.t.pradipto@gmail.com}
\affiliation{Consiglio Nazionale delle Ricerche CNR-SPIN, L'Aquila, Italy}
\author{M. Moretti Sala}
\affiliation{European Synchrotron Radiation Facility, CS 40220, 38043 Grenoble Cedex 9, France}
\author{L. C. Chapon}
\affiliation{Institut Laue Langevin, CS 20156, 38042 Grenoble Cedex 9, France}
\author{V. Simonet}
\affiliation{CNRS, Institut N\'eel, 38042 Grenoble, France}
\affiliation{Univ. Grenoble Alpes, Institut N\'eel, 38042 Grenoble, France}
\author{S. Picozzi}
\affiliation{Consiglio Nazionale delle Ricerche CNR-SPIN, L'Aquila, Italy}
\author{P. Lejay}
\affiliation{CNRS, Institut N\'eel, 38042 Grenoble, France}
\affiliation{Univ. Grenoble Alpes, Institut N\'eel, 38042 Grenoble, France}
\author{S. Petit}
\affiliation{Laboratoire L\'eon Brillouin, CEA, CNRS, Univ. Paris Saclay, CEA Saclay, F-91191 Gif-sur-Yvette, France}
\author{R. Ballou}
\affiliation{CNRS, Institut N\'eel, 38042 Grenoble, France}
\affiliation{Univ. Grenoble Alpes, Institut N\'eel, 38042 Grenoble, France}
\date{\today}

\begin{abstract}
We report our investigation of the electronic and magnetic excitations of \SNIO\ by resonant inelastic x-ray scattering at the Ir L$_3$ edge. The intra-\tg\ electronic transitions are analyzed using an atomic model, including spin-orbit coupling and trigonal distortion of the IrO$_6$ octahedron, confronted to {\it ab initio} quantum chemistry calculations. The Ir spin-orbital entanglement is quantified and its implication on the magnetic properties, in particular in inducing highly anisotropic magnetic interactions, is highlighted. These are included in the spin-wave model proposed to account for the dispersionless magnetic excitation that we observe at 90 \meV. By counterbalancing the strong Ni$^{2+}$ easy-plane anisotropy that manifests itself at high temperature, the anisotropy of the interactions finally leads to the remarkable easy-axis magnetism reported in this material at low temperature.
\end{abstract}


\maketitle

\begin{figure*}
	\resizebox{18cm}{!}{\includegraphics{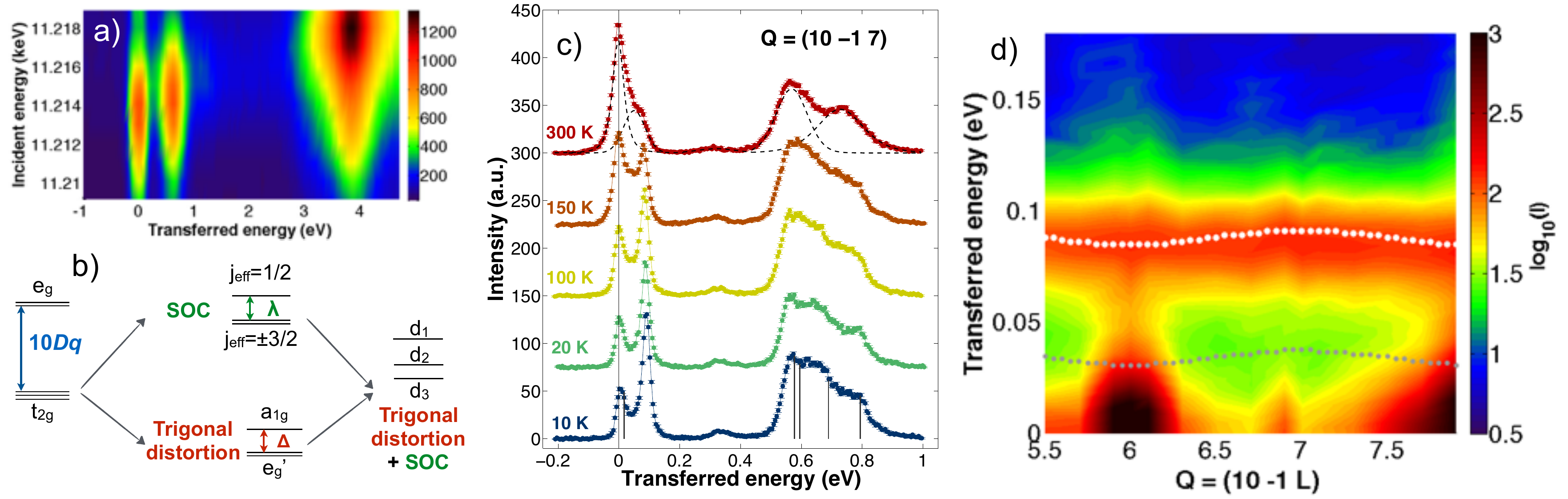}}
	\caption{\co (a) Incident energy dependence of the \SNIO\ RIXS spectra at the $L_3$ Ir edge measured at 300\K, showing the resonance of the \tg\ and \eg\ levels. (b) Splitting of the $5d$ levels due to the octahedral environment (10$Dq$), to its trigonal distortion and to SOC. (c) RIXS spectra at the reciprocal space position \textbf{Q}~=~(10~-1~7) between 10 and 300\K. The scans are offset by 75 counts for clarity. The dashed curves on the 300\K\ data correspond to least-square refinements of the excitations using Pearson-VII functions. The vertical lines on the 10\K\ curve corresponds to the eigenvalues of $\mathcal{H}$ calculated for $\lambda$=396\meV, $\Delta$=294\meV, $\alpha_{x,y}$=0 and $\alpha_z$= 0.1\eV. (d) RIXS map measured at 10\K\ along the (10 -1 {\bf}L) direction. The dotted lines show the spin-wave excitations calculated with $J_{xx} = J_{yy}$ = 20\meV, $J_{zz}$=46\meV\ and $D$=9\meV. The Ir and Ni main contributions are in white and grey respectively.}
	\label{fig:rixs}
\end{figure*}

Since the discovery of spin-orbit induced Mott insulator in Sr$_2$IrO$_4$ \cite{BJKim2008}, it has become apparent that electron correlation effects can be important in 5$d$ transition metal oxides when combined with spin-orbit coupling (SOC), despite rather wide $d$ electronic bands. In a perfect octahedral environment of $5d^5$ low-spin systems, e.g. in Ir$^{4+}$ oxide compounds, SOC splits the $t_{2g}$ band occupied by the 5 electrons, into a fully-occupied $j_{\rm eff}=\frac{3}{2}$ band and a narrow singly-occupied $j_{\rm eff}=\frac{1}{2}$ band. This band can be gapped by a fair amount of Hubbard repulsion driving the system toward an insulating state. 
This can persist even when the ideal $j_{\rm eff}=\frac{1}{2}$ state is not realized, i.e. when it is mixed with the $j_{\rm eff}=\frac{3}{2}$ states due to the Ir$^{4+}$ non regular octahedral environment (see Fig.~\ref{fig:rixs}(b)), as it is generally the case in the known iridates \cite{Moretti2014b,Gretarsson2013,Hozoi2014,Moretti2014}. The influence of SOC is not limited to transport properties. Another very interesting perspective in the field of iridates is the strong anisotropy of the magnetic interactions produced by spin-orbital entanglement. This is expected to give rise to novel exotic magnetic states and excitations, still to be discovered \cite{Witczak2014,Jackeli2009}.

In this context, the compounds of the family A$_3$MM'O$_6$ (A = alkaline-earth metal, M,M' = transition metal) are of interest because the M' site can be occupied by Ir$^{4+}$. Most members of the family crystallize in the K$_4$CdCl$_6$-derived rhombohedral structure in which alternating distorted trigonal prismatic MO$_6$ and nearly octahedral M'O$_6$ coordinations constitute chains along the rhombohedral axis (see Fig.~\ref{fig:Chi}) arranged on a triangular lattice. They show complex magnetic behaviors attributed mainly to frustration and low-dimensionality, enhanced by strong magnetocrystalline anisotropy and the presence of two magnetic species. \SNIO\ displays particularly intriguing properties: a complex antiferromagnetic structure below a N\'eel temperature T$_N$ of 70\K\ with slow spin dynamics and two consecutive correlated magnetic regimes when lowering the temperature \cite{Nguyen1995,Flahaut2003,Mikhailova2012,Lefrancois2014}. The ordered magnetic moments are aligned along the $c$-axis, but there is currently no explanation for the origin of the large anisotropy (single ion effect or exchange-driven mechanism) and its relation with the very high coercive field \cite{Singleton2014} of this material. The role played by SOC and anisotropic exchange is therefore a key issue.  Three groups have performed {\it ab initio} electronic structure calculations \cite{Sarkar2010,Zhang2010,Ou2014} confirming \SNIO\ Mott insulating state. Nevertheless, they disagree about the nature of the frontiers $5d$ electronic levels, the sign of the nearest-neighbor Ni-Ir interactions, and the magnetocrystalline anisotropy. The role of SOC has been investigated in the related Cu compound by resonant inelastic x-ray scattering (RIXS) \cite{Hill2012,Hill2013}.
However, the distortion of the Ir$^{4+}$ octahedral environment in Sr$_3$CuIrO$_6$ is monoclinic, which leads to nonequivalent Ir-O distances and O-Ir-O angles. The resulting $j_{\rm eff}=\frac{1}{2}$ and $j_{\rm eff}=\frac{3}{2}$ mixed states and their overlap with the Cu $3d$ orbitals are then at the origin of an unusual ferromagnetic exchange anisotropy arising from antiferromagnetic superexchange \cite{Hill2013}. Ir$^{4+}$ octahedra in \SNIO, instead, only experience a pure elongation along the trigonal $c$-axis, which preserves equivalent Ir-O distances. \SNIO\ is accordingly a nice candidate to study the influence of simple distortion on the electronic and magnetic properties.

Hereafter, we report RIXS experiment on \SNIO\, probing simultaneously its electronic and magnetic excitations. The local atomic effective Hamiltonian of \SNIO\ is derived, complemented by quantum chemistry {\it ab initio} calculations. The analysis of the magnetic excitations reveals a large Ising anisotropy in the Ni-Ir magnetic exchange which competes with a sizable single-ion planar anisotropy for the Ni ions. This competition explains the directional crossover of the magnetization observed at high temperature.

RIXS measurements were performed at the ID20 beam line of the European Synchrotron Radiation Facility. The beam line is equipped with a 2~m arm spectrometer, based on spherical Si(844) diced crystal analyzers. The combination of the Si(111) double-crystal monochromator with a channel-cut allows for a flexible choice of the energy-resolution down to 25\meV\ \cite{Moretti2013} at the Ir $L_3$ edge. RIXS has proven an ideal tool in the study of the iridate electronic structure \cite{Kim2012,Kim2012b,Hill2012,Gretarsson2013}, as it directly probes $5d$ states via two successive dipole transitions \cite{Ament2011,Moretti2014c}. Measurements were performed between 300 and 10\K\ on a single crystal of \SNIO, grown using the flux method \cite{Lefrancois2014}. Fig.~\ref{fig:rixs}(a) shows a RIXS map obtained from inelastic scans at different incident photon energies. The resonance of the Ir$^{4+}$ electronic levels is observed at 11.214\keV\ for the \tg\, and at 11.218\keV\ for the \eg. For all measurements presented below, the incident photon energy was fixed to 11.214\keV, which enhances the excitations within the \tg\ manifold.

The RIXS spectra of \SNIO\ measured at the momentum transfer {\bf Q}~=~(10~-1~7) of a forbidden Bragg peak position are shown Fig.~\ref{fig:rixs}(c). We first focus on the room temperature measurement where four features are visible beside the elastic peak, at the energies of 50(5), 322(20), 568(2) and 728(5)\meV, none of them showing any detectable dispersion within the lowest instrumental resolution. The most intense peaks at 568(2) and 728(5)\meV\ are interpreted as $d$-$d$ excitations within the $t_{2g}$ levels \cite{Hill2012,Hill2013,Hozoi2014}, i.e. as transitions of the hole from the ground state to the two lower lying filled states (see Fig.~\ref{fig:rixs}(b)). In addition to the SOC splitting of the \tg, a trigonal crystal field indeed further splits the $j_{\rm eff}=\frac32$ ground manifold. This effect can be quantified using a local atomic model for the single hole in the \Tg\ states with the Hamiltonian: 

\begin{multline}
\mathcal{H} =  \Delta(\frac{2}{3}|a_{1g}\rangle \langle a_{1g}| - \frac{1}{3}|e_g^{\prime\pm}\rangle\langle e_g^{\prime\pm}|) + \lambda{\bf L\cdot S}~- \sum_{i \in {x,y,z}}\alpha_i S_i 
\end{multline}
where $\Delta$ describes the \tg\ orbitals splitting due to the IrO$_6$ octahedron trigonal distortion ($\Delta>0$ for an axial elongation) and $\lambda$ is the SOC parameter. The symmetry-adapted $|a_{1g}\rangle$ and $|e_{g}^{\prime\pm}\rangle$ wave-functions for trigonal symmetry are defined in the Sup. Mat. \cite{SupMat}. The $10Dq$ splitting between \tg\ and \eg\ levels is of the order of 4\eV\ (see Fig.~\ref{fig:rixs}(a)), i.e. much larger than $\lambda$ and $\Delta$, and therefore the \eg\ levels can safely be ignored. The last terms, $\alpha_{i}S_{i}=-2J_{ii}<S^{Ni}_i>S_i$ ($i=x,y,z$),
are the molecular field components produced on the Ir spin by its two Ni$^{2+}$ nearest neighbors. For the analysis of the 300\K\ data, far above T$_N$, we first assume $\alpha_{x,y,z}=0$, i.e. weak influence of the magnetic interactions.

\begin{figure}[t]
	\centering
	\resizebox{8cm}{!}{\includegraphics{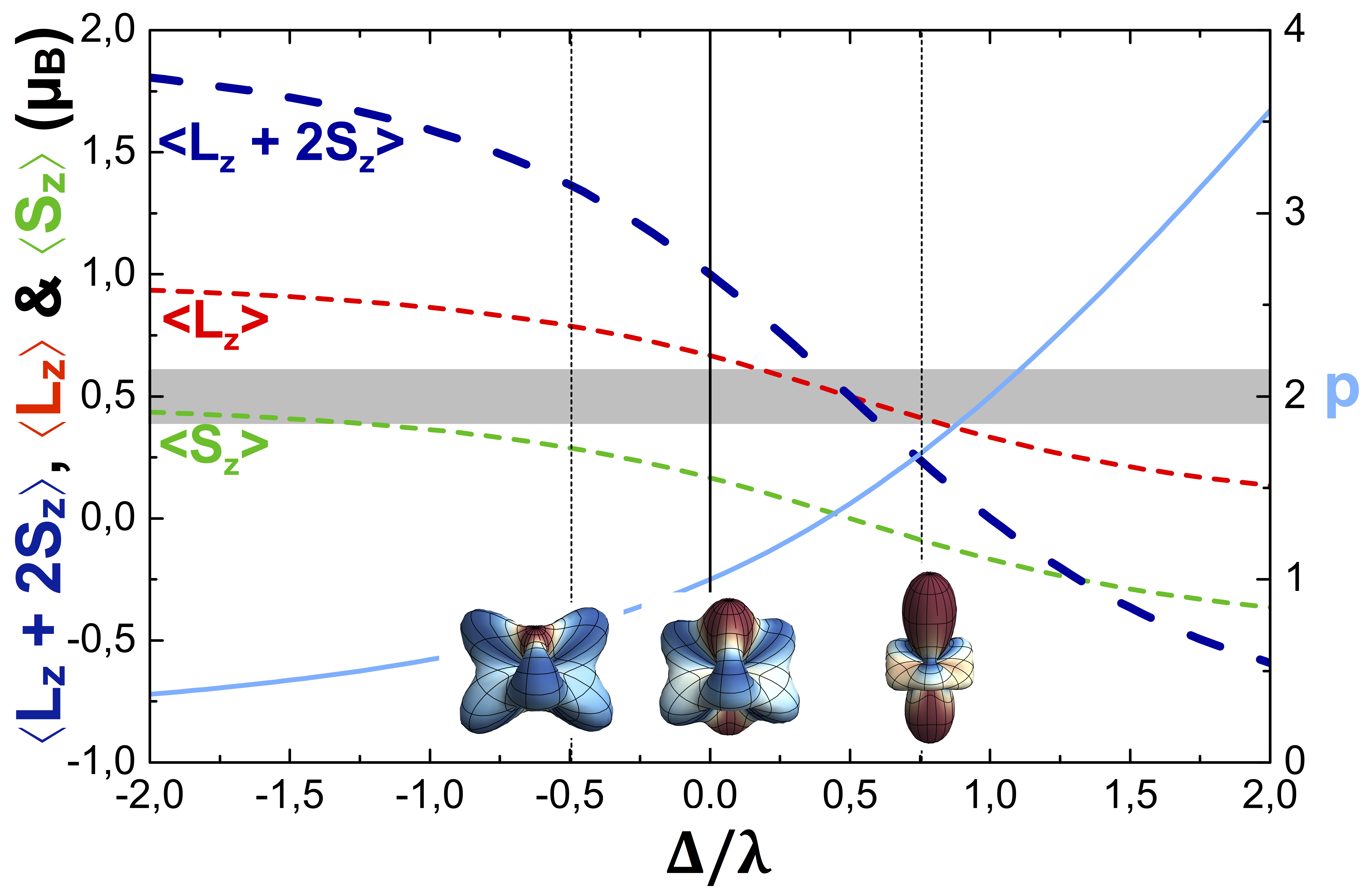}}
	\caption{\co Evolution of the $z$ components of the orbital <L$_z$>, spin <S$_z$>, total magnetic moment $<L_z+2S_z>$ and of $p$ as a function of $\frac\Delta\lambda$. The grey stripe shows the Ir magnetic moment refined from neutron diffraction \cite{Lefrancois2014}. The amplitudes of the probability of presence of the \tg\ hole are shown at the bottom: for a perfect octahedral environment ($\frac\Delta\lambda=0$), for a trigonal compression ($\frac\Delta\lambda=-0.52$) and for a trigonal elongation ($\frac\Delta\lambda = 0.73$) of the octahedron. The latter two are in agreement with the RIXS results but only the positive $\frac\Delta\lambda$ is compatible with quantum chemistry calculations. The surfaces of constant amplitude of probability are colored from red for pure $|a_{1g},\uparrow\rangle$ state, to blue for pure ($i|e_g^{\prime+},\downarrow\rangle + |e_g^{\prime-},\downarrow\rangle$) state, through white for an equal mix of these states.}
	\label{fig:GS}
\end{figure}

\begin{figure}[h]
	\centering
	\resizebox{8cm}{!}{\includegraphics{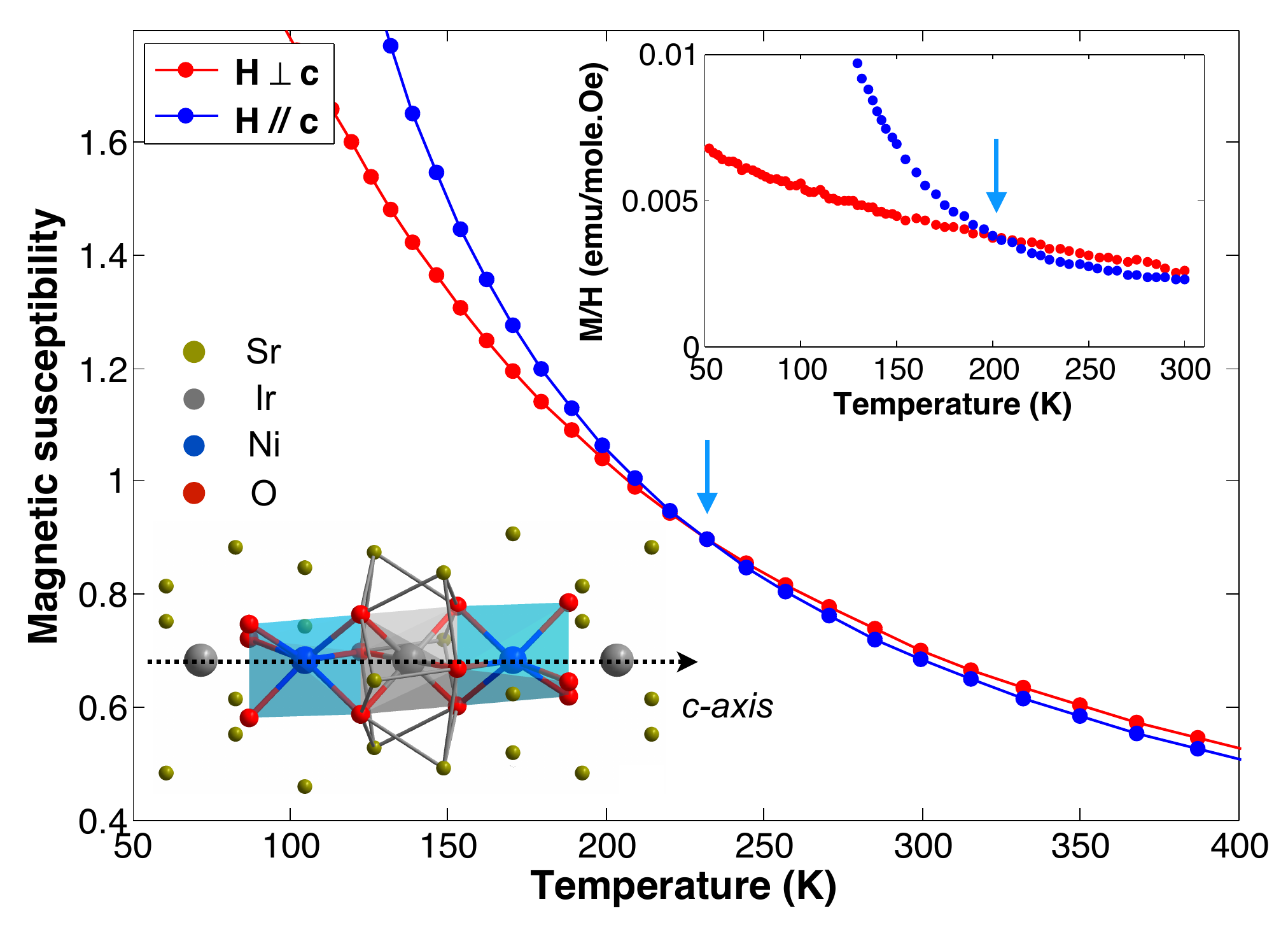}}
	\caption{\co Magnetic susceptibility of \SNIO\ calculated by Monte Carlo simulation using the spin-wave parameters. The blue (red) curve is obtained for the magnetic field applied along (perpendicular to) the chain direction. Top insert: Comparison with the measurements reported in Ref. \cite{Lefrancois2014}. The arrows indicate the crossing of the curves. Bottom insert: Detail of the \SNIO\ structure with the central distorted IrO$_6$ octahedron along the chain axis.}
	\label{fig:Chi}
\end{figure}

The values of $\lambda$ and $\Delta$ are obtained by constraining the eigenvalues of $\mathcal{H}$ to the energies of the \tg\ manifold excitations determined from the experiment. Two solutions exist, depending on the sign of $\Delta$: $\lambda$=396(1)\meV\ and $\Delta$=294(7)\meV\ for the first one and $\lambda$=417(4)\meV\ and $\Delta$=-218(8)\meV\ for the second one. The wave-functions (shown in Fig.~\ref{fig:GS}) for the highest doublet in the \Tg\ subspace (occupied by the hole) are:
\begin{equation}
	\left\{
	\begin{array}{l}
	|0,\uparrow\rangle = \frac{ip|a_{1g},\uparrow\rangle + i |e_g^{\prime+},\downarrow\rangle + |e_g^{\prime-},\downarrow\rangle}{\sqrt{p^2 + 2}}\\[0.5cm]
	|0,\downarrow\rangle = \frac{p|a_{1g},\downarrow\rangle - |e_g^{\prime+},\uparrow\rangle - i|e_g^{\prime-},\uparrow\rangle}{\sqrt{p^2 + 2}}
	\end{array}
	\right.
	\label{eq:gs}
\end{equation}
with the evolution of $p$ with $\Delta/\lambda$ shown in Fig.~\ref{fig:GS}.
The sign of $\Delta$ can not be determined by the RIXS analysis alone. However, the Ir magnetic moment refined in neutron diffraction \cite{Lefrancois2014} seems in better agreement with the $z$ component of the total magnetic moment, $<L_z+2S_z>$, calculated for a positive $\Delta$ (0.25~$\mu_B$) than for a negative one (1.35~$\mu_B$), as shown in Fig.~\ref{fig:GS}. 
Nevertheless, a strong reduction of the magnetic moment by quantum fluctuations, enhanced by low dimensionality or geometric frustration, is not excluded.

\par 
Another argument in favor of $\Delta>0$ comes from {\it ab initio} calculations. For these, we used a quantum chemistry inherited computational scheme \cite{Hill2012,Katukuri2012,Gretarsson2013,Hozoi2014}, in which it was shown that the combination of Multireference Configuration Interaction method with SOC (MRCI+SOC) is able to reproduce the RIXS spectra. The local Ir-$5d$ electronic structure has been investigated in the basis of multi-configurational wave-function-based methods \cite{SupMat} which is implemented in MOLCAS 7.8 code \cite{Aquilante2010} to ensure a proper description of the multiplet physics. 

\begin{table}[b]
\begin{center}
\begin{threeparttable}
\begin{tabular}{lccccccccccc}
\toprule[1pt]
& \multicolumn{3}{c}{No SOC} && \multicolumn{3}{c}{With SOC} && \multicolumn{3}{c}{Exp. values} \\
\cmidrule[0.5pt]{2-4}\cmidrule[0.5pt]{6-8}\cmidrule[0.5pt]{10-12}
Configuration & $\epsilon_0$ & $\epsilon_1$ & $\epsilon_2$ && $\epsilon_0$ & $\epsilon_1$  & $\epsilon_2$ && $\epsilon_0$ & $\epsilon_1$  & $\epsilon_2$ \\ 
\cmidrule[0.5pt]{2-4}\cmidrule[0.5pt]{6-8}\cmidrule[0.5pt]{10-12}
& 0.00 & 0.21 & 0.21 && 0.00 & 0.64 &0.80 && 0.00 & 0.57 &0.73 \\
 \midrule[0.5pt]
$\vert\psi_1\rangle=d_1^1d_2^2d_3^2$ & 1.00 &         &         && 0.46 & 0.54 & 0.00 && 0.58 & 0.42 & 0.00 \\
$\vert\psi_2\rangle=d_1^2d_2^1d_3^2$ &         & 1.00 &         && 0.27 & 0.23 & 0.50 && 0.21 & 0.29 & 0.50 \\
$\vert\psi_3\rangle=d_1^2d_2^2d_3^1$ &         &         & 1.00 && 0.27 & 0.23 & 0.50 && 0.21 & 0.29 & 0.50 \\
\bottomrule[1pt]
\end{tabular}
\end{threeparttable}
\caption{SD-MRCI results for the $t_{2g}$ level splittings (in\eV) and
weights of the different Ir-$t_{2g}$ configurations. In the absence of SOC, $\epsilon_1 = \epsilon_2$ corresponds to the energy of the doubly degenerate $E_g'$ state. The energies and weights extracted from the experiment for $\Delta$>0 are shown in the last column.
}
\label{calc}
\end{center}
\end{table}

The 54 spin-orbit Ni-Ir coupled states and their energies were calculated for the cluster shown in Fig.~\ref{fig:Chi} \cite{SupMat}. In order to estimate the excitations only within the Ir-$5d$ orbitals, we performed calculations by replacing the Ni$^{2+}$ with non-magnetic Zn$^{2+}$. Without SOC, we observe three doublet states, in which the degenerate $^2E'_g$ doublets lie 0.21\eV\ higher than the ground state (see Table~\ref{calc}). This is in perfect agreement with the energy windows calculated with Ni$^{2+}$, confirming that these correspond to the excitation energies within the Ir-5$t_{2g}$ manifold. Furthermore, the ground doublet state corresponds to the $d_1^1d_2^2d_3^2$ configuration, where $d_1$ is the $a_{1g}$ orbital that can be written as $d_{z^2}$, while $d_2$ and $d_3$ correspond to the $e_g'$ orbitals in the trigonal representation. One can hence conclude that the $a_{1g}$ orbital lies higher in energy than the $e_g'$ ones, in accordance with a positive $\Delta$.

Taking into account the SOC, two excitations are calculated at 0.64 and 0.80\eV. These values are remarkably close to the energies found in the RIXS spectra, confirming that the associated peaks correspond to the splitting of the $j_{\rm eff}=\frac{3}{2}$ quartet into two doublets. The weights of different $t_{2g}^5$ contributions to the spin-orbit coupled doublet states are shown in Table \ref{calc}. 
The ground state is not an equal mixture of the $t_{2g}$ orbitals, the weight of $\vert \psi_1\rangle$ being larger than the other two degenerate configurations, with a ratio of $0.46:0.27:0.27$ (to be compared with $0.33:0.33:0.33$ in a perfect cubic environment). Overall, the room temperature RIXS measurements confronted to the {\it ab initio} calculations are fully consistent and allow us to describe quantitatively the electronic spectrum of the Ir$^{4+}$ and its departure from the perfect $j_{\rm eff}=\frac{1}{2}$ state due to its non-cubic environment. An Ir-5$t_{2g}$ orbitals splitting $\Delta$ ranging between 0.21 and 0.30\eV\ is obtained, leading to a ratio $\Delta/\lambda \approx 0.53-0.74$ ($\lambda$=0.396\eV\ estimated from RIXS). This scheme does not explain the weak signal observed at 322\meV. Its origin is unclear, but a feature with similar energy and spectral weight was reported in almost all the iridates \cite{Hill2012,Gretarsson2013,Hozoi2014}, suggesting that it might be an intrinsic characteristics of this class of materials.

Lowering the temperature down to 10\K, the two intense peaks related to the $t_{2g}$ electronic transitions change shape and acquire a substructure (see Fig.~\ref{fig:rixs}(c)). This reflects the increasing influence, as the temperature decreases, of the Ni-Ir nearest neighbor magnetic exchange interactions on the Ir electronic states. The molecular field produces a splitting of the peaks that can be calculated from the last term of Eq. 1.  Although the features observed in the range 500-900\meV\ do not allow to univocally determine the values of $\alpha_{x,y,z}$, an example of splitting is shown on the 10 K spectrum of figure \ref{fig:rixs}(c), in reasonable agreement with the experiment. It is obtained with $\alpha_{x,y}$=0 and $\alpha_z$=0.1 eV, agreeing with the strong uniaxial character of the magnetic structure, and with the value of the $J_{zz}$ interaction determined hereafter.

At 300\K, the excitation at the lowest energy is broad with a position of the maximum at $\approx$50(5)\meV. It rises in intensity while narrowing as the temperature is lowered (see Fig.~\ref{fig:rixs}(c)), and its energy rapidly shifts to $\approx$90(5)\meV\ at 100 K, below which it remains constant. This excitation cannot be accounted for by the local electronic level scheme derived previously. It is attributed to a magnetic excitation whose observation by RIXS is allowed in this system \cite{SupMat}. The absence of any detectable dispersion is illustrated by the RIXS measurement along the (10 -1 L) direction (see Fig.~\ref{fig:rixs}(d)). Another magnetic excitation was identified at 35\meV\ by inelastic neutron scattering (INS) on a \SNIO\ powder sample \cite{Wu2015}. It was also shown to persist up to 200\K, i.e much above T$_N$, a behavior attributed to the presence of strong spin correlations associated to the low-dimensional nature of the magnetism. We succeeded in reproducing the spin-wave modes at 35 and 90\meV\ observed by INS and RIXS respectively using the Holstein-Primakov formalism within the linear approximation (see Fig.~\ref{fig:rixs}(d)) \cite{SupMat}. We used a simple model, in line with the reported magnetic structure \cite{Lefrancois2014,SupMat}, of isolated chains with (i) strong anisotropic antiferromagnetic exchange interactions between the nearest-neighbor Ni and Ir ions along the chain: $J_{xx}$=$J_{yy}$=20(2)\meV, $J_{zz}$=46(2)\meV\, and (ii) magnetocrystalline anisotropy of Ni$^{2+}$: $DS_z^2$ term with $D$=9(1)\meV\ corresponding to an easy-plane single-ion anisotropy perpendicular to the chain axis. 
The effective interchain interaction was estimated from both the ordering temperature and the coercive field. It is more than one order of magnitude smaller than the intrachain interactions, affecting negligibly the spin excitations. It is furthermore irrelevant to the spin wave dispersion along the chains. It can thus be safely ignored \cite{SupMat}.
Extracting the ion-dependent neutron spectral weight of the excitations clearly demonstrates that Ir(Ni) mainly contributes to the high(low) mode \cite{SupMat}.This explains the absence of the INS 35\meV\ mode in the RIXS data since the measurements at the Ir $L_3$ edge are mainly sensitive to the resonant inelastic signal of Ir. Our model is highly constrained by the energies and dispersion of the modes extracted from the RIXS and INS data \cite{SupMat} and can uniquely account for the experimental results if one considers this easy-plane single-ion term. This term is a hallmark difference between the physics of \SNIO\ and Sr$_3$CuIrO$_6$, which otherwise shows rather similar anisotropic exchanges \cite{Hill2013}.
The strength of the exchange anisotropy is driven by $p$, which is related to the relative weight of the $|a_{1g}\rangle$ and $|e_g^{\prime}\rangle$ wave functions in Eq.\ref{eq:gs}. Those are respectively preserving or not L$_z$ when flipping S$_z$, hence contributing to the isotropic (resp. anisotropic) part of the interactions \cite{Hill2013}. The parameter $p$ varies with the octahedral distortion $\Delta$ and might be tunable by a chemical or external pressure.  

To further support this model, it is found that the Ni anisotropy perfectly matches the one reported for the isostructural compound Sr$_3$NiPtO$_6$, for which $D\approx9$\meV\ \cite{Chattopadhyay2010} and which is even considered as a prototypal example of a "large-$D$" system. Furthermore, the competition between strong exchange anisotropy ($J_{zz}/J_{xx}$) and single-ion anisotropy ($DS_z^2$) on the Ni site produces a hallmark feature in the magnetization measurement: at high temperature, the susceptibility measured for a magnetic field applied along the c-direction is lower than the one obtained with a field in-plane. Below 200\K, the two curves cross due to the increasing influence of the large exchange anisotropy when correlations build-up, ultimately leading to a uniaxial antiferromagnetic order. The directional crossover in the magnetic susceptibility is nicely reproduced by Monte-Carlo simulations using a one-dimensional chain model and the parameters extracted from the spin-wave calculations (see Fig.~\ref{fig:Chi}) \cite{SupMat}. Finally, the large coercive field reported in \SNIO\ is simply explained as the minimum field responsible for the coherent reversal of the spin chains, its value agreeing with the interchain coupling \cite{SupMat}.

To summarize, we have determined the $5d$ Ir electronic scheme in the \SNIO\ chain compound through our RIXS measurements and their analysis. The resulting spin-orbital entangled ground state produces a strong anisotropy of the interactions, which is evidenced in our analysis of the spin-wave excitations including the two gapped modes seen by RIXS and neutron scattering. This uniaxial anisotropy of the interactions competes with a strong Ni$^{2+}$ single ion easy-plane anisotropy, a unique feature of this system in the series. Comparing our magnetic Hamiltonian to the one of a recent numerical study of a mixed-spin XXZ chain with single-ion anisotropy, it is found that, in the parameter space of exchange and single-ion anisotropies, \SNIO\ lies close to the boundary separating the observed ordered Ising ferrimagnetic phase from a disordered XY phase \cite{Qiang2015}. These results demonstrate the potential of SOC in triggering emerging physics. In mixed 3d-5d transition metal chain systems, it can allow to probe unexplored regions of their phase diagram and quantum phase transitions.

\section*{Acknowledgments}

We thank A. Al-Zein for this help during the RIXS experiment and A. Hadj-Azzem for his collaboration on the crystal synthesis. 
A.-M. P. and S. P. thank R. Broer for providing with the computational facilities in the local computer cluster at the University of Groningen.

	

\widetext
\pagebreak
\begin{center}
	\textbf{\large Supplemental Materials: Anisotropic interactions opposing magnetocrystalline anisotropy  in \SNIO.}
\end{center}
\setcounter{equation}{0}
\setcounter{figure}{0}
\setcounter{table}{0}
\setcounter{page}{1}
\makeatletter
\renewcommand{\theequation}{E\arabic{equation}}
\renewcommand{\thefigure}{F\arabic{figure}}
\renewcommand{\bibnumfmt}[1]{[R#1]}
\renewcommand{\citenumfont}[1]{R#1}

\section{Definitions of the orbitals used in the simple Hamiltonian accounting for the RIXS measurements} \label{app:orb} 

\begin{equation}
\left\{
\begin{array}{ccl}
(xy) & = & -\frac{\imath}{\sqrt{2}}(|2,2\rangle-|2,-2\rangle)\\
(yz) & = & \frac{\imath}{\sqrt{2}}(|2,1\rangle+|2,-1\rangle)\\
(zx) & = & -\frac{1}{\sqrt{2}}(|2,1\rangle-|2,-1\rangle)\\
(x^2-y^2) & = & \frac{1}{\sqrt{2}}(|2,2\rangle+|2,-2\rangle)\\
(z^2) & = & |2,0\rangle
\end{array}
\right.
\label{eq:xyz}
\end{equation}

\begin{equation}
\left\{
\begin{array}{ccl}
|a_{1g}\rangle & = & (z^2)\\
|e_g^{\prime+}\rangle & = & \sqrt{\frac{2}{3}}(x^2-y^2)-\sqrt{\frac{1}{3}}(zx)\\
|e_g^{\prime-}\rangle & = & \sqrt{\frac{2}{3}}(xy)+\sqrt{\frac{1}{3}}(yz)\\
|e_g^+\rangle & = & \sqrt{\frac{1}{3}}(x^2-y^2)+\sqrt{\frac{2}{3}}(zx)\\
|e_g^-\rangle & = & \sqrt{\frac{1}{3}}(xy)-\sqrt{\frac{2}{3}}(yz)
\end{array}
\right.
\label{eq:t2g}
\end{equation}

\section{{\it Ab initio} calculation details}

The local Ir-$5d$ electronic structure has been investigated in the basis of multi-configurational Complete Active Space Self-Consistent Field (CASSCF) \cite{Roos1980} method which is implemented in MOLCAS 7.8 code \cite{Aquilante2010b} to ensure a proper description of the multiplet physics. To make such multi-electron approach feasible, a small part representing the crystal in which a finite number of atoms are treated explicitly in the calculation, called a `cluster', has to be chosen. In our case, we include an IrO$_6$ octahedra, its neighboring NiO$_6$ trigonal prismatic coordinations and the adjacent Sr atoms (see insert of Fig. 2 in the article). This choice is expected to provide an accurate description of the charge distribution in the region of IrO$_6$ octahedra. To model the crystalline environment, we employ a standard embedding procedure of the cluster \cite{Sadoc2007, Maurice2012} including (i) a set of optimized point charges \cite{Roos1969}, and (ii) the \textit{Ab Initio} Embedding Model Potentials \cite{Barandiaran1988} to avoid artificial excessive polarization of the cluster electrons towards the point charges. An active orbital space has been chosen to include the $t_{2g}$-like orbitals of the Ir-$5d$ shell and the magnetic orbitals of the Ni atoms to describe the excitations within the Ir-$5t_{2g}$ manifold. The chosen active space is enlarged to include also the Ir-$5e_g$ orbitals when the $t_{2g}\rightarrow e_g$ excitations are to be considered. The dynamical correlation is then taken into account at a Single and Double Multireference Configuration Interaction (SD-MRCI) level using the Restricted Active Space Self-Consistent Field \cite{Malmqvist1990} module of MOLCAS, in which a maximum number of 2 electrons are allowed to be excited from the $2p$ orbitals of oxygen atoms surrounding the iridium, without relaxing the orbitals from CASSCF. Further enlarging the external active space in the SD-MRCI step, e.g. to include all doubly occupied Ni-$3d$ shells or the empty $t_{2g}$ orbitals from the Ir-6$d$ shell to improve the electron correlation description, does not significantly change the results. The scalar relativistic effects are accounted for with the Douglass-Kroll-Hess Hamiltonian \cite{Douglas1974,Hess1986}. The wavefunctions and energies of the spin-orbit free multiplets obtained at the SD-MRCI level are then included for the spin-orbit coupling (SOC) calculation using the Spin-Orbit Restricted Active Space State Interaction method \cite{Malmqvist1989, Malmqvist2002} as implemented in MOLCAS code. The cluster `molecular' orbitals are expanded with ANO-RCC basis sets \cite{Roos2005} in all calculations, with the contraction schemes ($8s7p5d2f$), ($6s5p4d$), ($4s3p1d$), and ($6s5p2d1f$) for the Ir, Ni, O, and Sr atoms, respectively. \\

As expected, due to a large crystal field splitting 10$Dq$ of about 3.7\eV\ (estimated from Single and Double MRCI + SOC, in good agreement with the experimental value of 4\eV), all 5 electrons of Ir-$5d$ are arranged in a low-spin $S=\frac{1}{2}$ manner to occupy the $t_{2g}$-like orbitals. In the spin-orbit free framework, these three doublet states of Ir$^{4+}$ and the $S=1$ states of the two Ni$^{2+}$ ions can couple to produce three sets of one sextet, two quartet and two doublet states, lying in energy windows of $0-33$\meV\ and $0.24-0.26$\eV. This gives an estimate to the Ir-5$t_{2g}$ splitting of $0.2-0.24$\eV\ due to the non-cubic crystal field. Allowing these multiplets to interact via SOC results in a total of 54 spin-orbit coupled states lying in energy windows of $0-22$\meV, $0.64-0.66$\eV, and $0.82-0.84$\eV. 
In order to get better estimate on the excitations only within the Ir-$5d$ orbitals, further calculations have been performed by replacing the Ni$^{2+}$ with non-magnetic Zn$^{2+}$ ions, as presented in the main article. 
One result of these SD-MRCI calculations is that the three lowest doublets are always dominated by $t_{2g}^5$ configurations, weighting more than 98\% of the total contributions, confirming a minor $t_{2g}-e_g$ mixing to the ground state wavefunction.

\section{Magnetic excitation and Spin waves calculations} \label{app:SW} 

We checked beforehand that single magnon excitations in \SNIO\ can effectively be detected by RIXS. The RIXS scattering intensity is given by the Kramers-Heisenberg formula \cite{rxs1994} and is proportional to the matrix elements of the polarization dependent dipole operator $C_\epsilon = \epsilon \vec{r}$ as:
\begin{equation}
I_{\hbar\omega \rightarrow \hbar\omega'} \propto |\langle 5d_f |C_{\epsilon'}|3p_{\frac32}\rangle\langle 3p_{\frac32}|C_\epsilon|5d_i\rangle|^2,
\label{eq:rixs_int}
\end{equation}
where $5d_i$ and $5d_f$ stand for the initial and final d states and $3p_\frac32$ for the intermediate states. Using the dipole matrix reported in ref~\cite{deGroot1998}, we can compute the matrix elements displayed in Eq.~\ref{eq:rixs_int} for the different possible intermediate $3p_{\frac32}$ states. They are gathered in Table I for initial and final $d$ states in the ground level with respect to $d$-$d$ excitations ({\it i.e.} $|0,\uparrow\rangle$ and $|0,\downarrow\rangle$ states defined in the main text). All of them are not null thereby implying that single magnons with spin deviations in the ground level do contribute to the scattering.\\

\begin{center}
	\begin{tabular}{c|cc}
		& $|\frac32,\pm\frac32\rangle$ & $|\frac32,\pm\frac12\rangle$ \\[.2cm]
		\hline
		$|0,\uparrow\rangle \rightleftharpoons|0,\downarrow \rangle$ & $\frac{i2\sqrt2 p}{2+p^2}$ & $ -\frac{i\sqrt2(2+5p+2p^2)}{3(2+p^2)}$  \\
	\end{tabular}
	\label{tab:rixs_int}
	\captionof{table}{Computed $\langle 0,\uparrow(\downarrow) |C_{\epsilon'}|3p_{\frac32}\rangle\langle 3p_{\frac32}|C_\epsilon|0,\downarrow(\uparrow)\rangle$ quantities for the different $3p_\frac32$ states.}
\end{center}

To model the spin dynamics, spin-wave calculations have been performed using the {\it Spinwave} software developed at Laboratoire L\'eon Brillouin for inelastic neutron scattering experiments. Based on the Holstein-Primakov approximation, the code diagonalises any bilinear spin Hamiltonian and takes into account (isotropic or anisotropic) exchange couplings acting between
neighboring spins, as well as single ions anisotropy terms.
The initial spin configuration is obtained from the mean field solution of the following Hamiltonian:
\begin{equation}
\mathcal{H} = \sum_i J_{xx}(S_i^x S_{i+1}^x + S_i^y S_{i+1}^y ) +  J_{zz} S_i^z S_{i+1}^z + \sum_{i \in Ni} D (S_i^z)^2
\end{equation}
which describes an isolated chain with alternating spin 1 and spin $\frac12$ antiferromagnetically coupled along the chains \cite{Lefrancois2014b}. The exchange tensor acting between the first neighbors along the {\it c} axis is highly anisotropic, defined by $J_{xx}$ and $J_{zz}$. $\mathcal{H}$ also takes into account the magnetocrystalline anisotropy of the Ni spins, modeled by $D (S_i^z)^2$, and which corresponds to an easy plane anisotropy perpendicular to the $z$ axis. The point symmetry between Ni and Ir site is 3, thus allowing for antisymmetric exchange interactions with a Dzyaloshinskii-Moriya D-vector pointing along the c-axis. The upper limit estimate for this interaction, $\approx5$\meV, shows that it is not the leading term in the Hamiltonian, and has thus been neglected, together with the interchain interactions (see below).\\

Figure~\ref{fig:sw} shows the calculations performed with the parameters: $J_{xx}$ = 20\meV, $J_{zz}$ = 46\meV\ and $D$ = 9\meV, along a $Q$-direction corresponding to the zone explored during the RIXS experiment (see left Fig.~\ref{fig:sw}). Our calculation shows the presence of two excitations around 35 and 90\meV\ with a small dispersion of the order of 10\meV. \\

We also calculated the powder average of the spin wave spectrum in order to compare our model with the powder neutron inelastic experiment performed bu Wu \etal\ \cite{Wu2015b} (see right Fig.~\ref{fig:sw}). Because of the integration over the momentum directions, the resulting spectra lack directional information but give a weighted density of states for each magnitude of Q. This results in two band-like excitations, the lower one reproducing fairly well the INS results. The values of the different parameters are highly constrained by the results from the RIXS and INS experiments. Indeed the position of the maxima of the two modes are equal to $J_{zz} - D$ and to $2J_{zz}$, whereas the dispersion is related to $\nicefrac{J_{xx}}{J_{zz}}$: The amplitude of the dispersion decreases as $J_{xx}$ decreases and as $J_{zz}$ increases.\\

\begin{figure*}[H]
	\begin{center}
		\includegraphics[scale=0.53,trim=0 -0.4cm 0 0cm,clip]{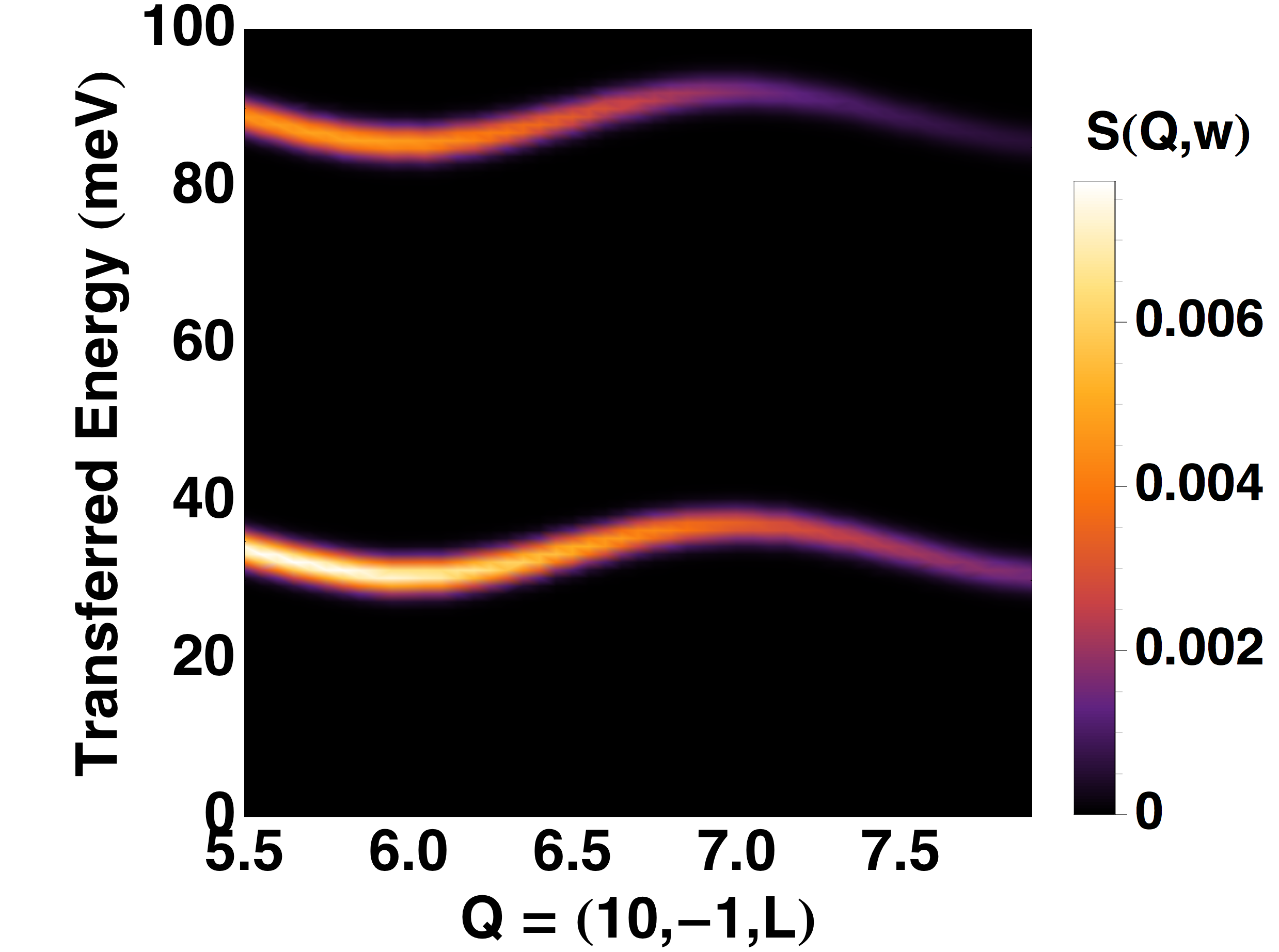}
		\includegraphics[scale=0.25,trim=0 0.9cm 0 0,clip]{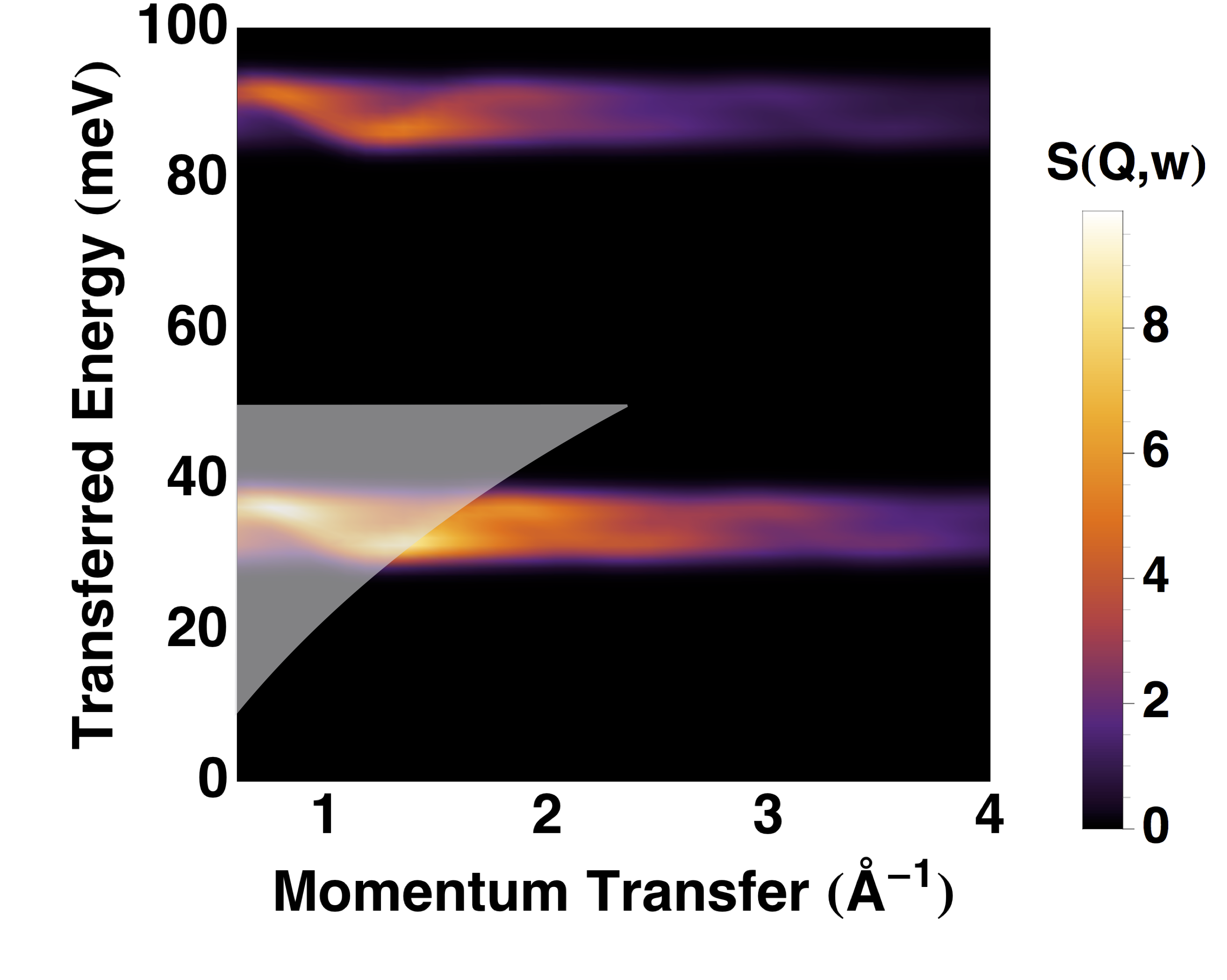}
		\caption{ \co Calculated spin waves with exchange interactions $J_{xx}$ = $J_{yy}$ = 20\meV\ and $J_{zz}$ = 46\meV\ and magnetocrystalline anisotropy $D$ = 9\meV\ for a single crystal (left) and for a polycrystalline sample (right). The white zone represents the region which was not accessible in the INS experiment from Fig.4 of ref. \cite{Wu2015} up to 50 meV.} 
		\label{fig:sw}
	\end{center}
\end{figure*}

\begin{figure*}[H]
	\begin{center}
		\includegraphics[scale=0.58,trim=0 0 4.5cm 0,clip]{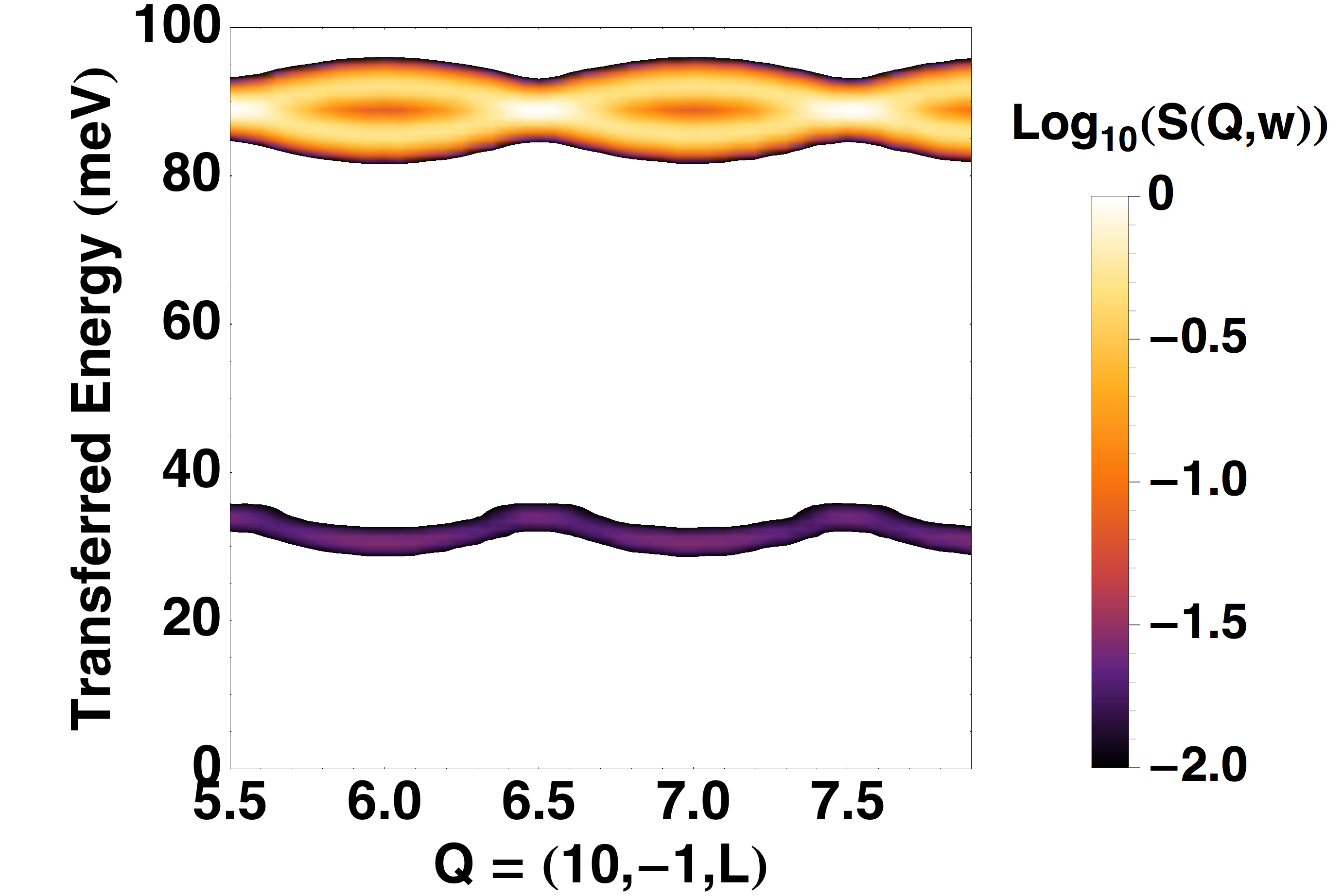}
		\hspace{0.5 cm}
		\includegraphics[scale=0.58,trim=1.8cm 0 0 0,clip]{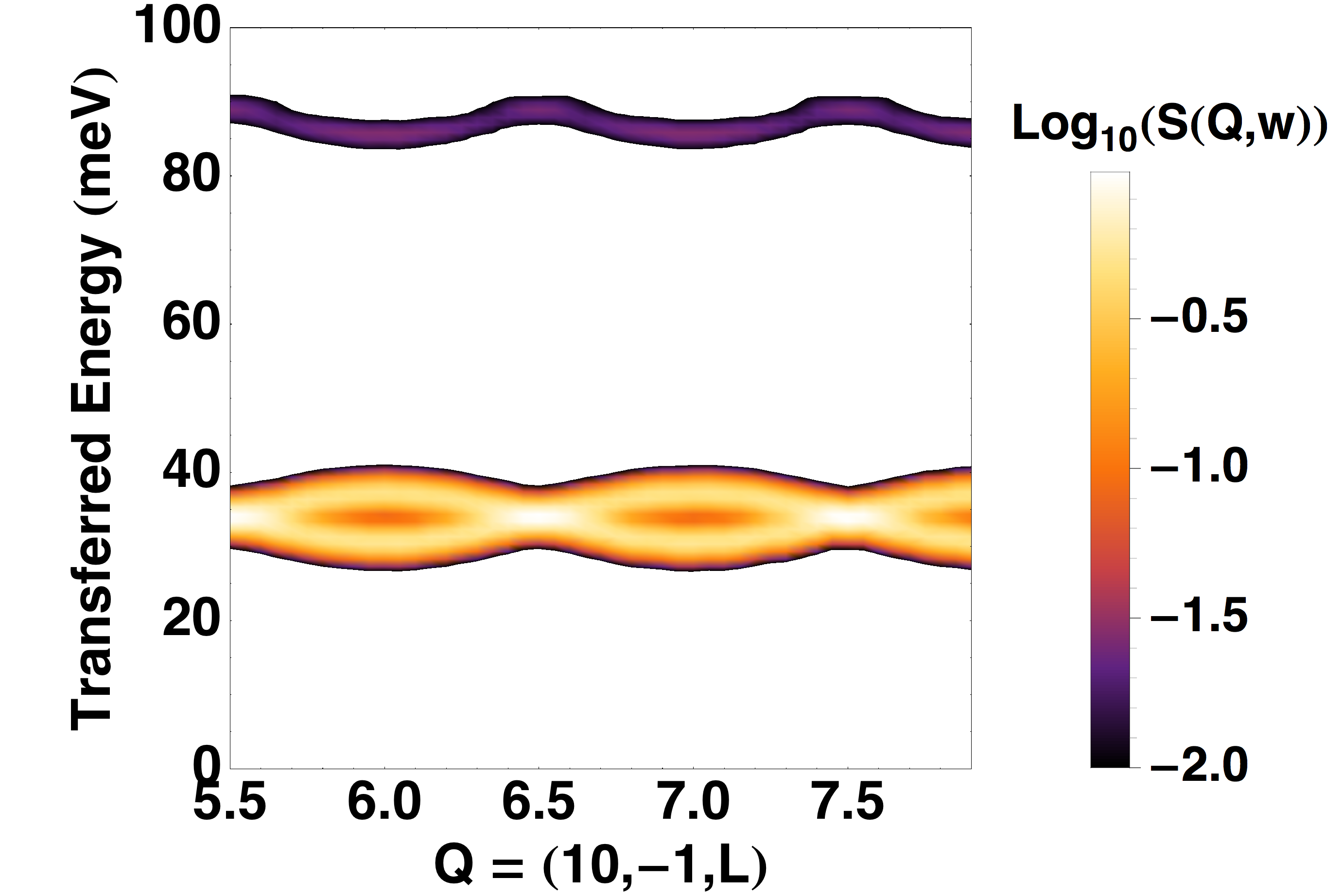}
		\caption{ \co Contributions of the Ir ions (left) and Ni ions (right) in the calculated spin waves with exchange interactions $J_{xx}$ = 20\meV\ and $J_{zz}$ = 46\meV\ and magnetocrystalline anisotropy $D$ = 9\meV. The colorbar is in log scale.} 
		\label{fig:sw_contr}
	\end{center}
\end{figure*}

The {\it Spinwave} software also allows one to identify the contributions arising from the different magnetic species. Indeed, since a spin-wave is a collective excitation, each spin within the unit cell contributes with however a different weight. A close look at those individual contributions shows that the lower branch is essentially due to the Ni contribution while the upper branch arises mainly from the Ir contribution (see Fig.~\ref{fig:sw_contr}). This explains why we did not observed the low-energy excitation in the RIXS experiment as the energy was fixed at the Ir-$L_3$ resonance edge. The reason why the high-energy excitation was not seen in INS is probably due to the weakness of the signal in the rather high $Q$-range investigated. Overall, the calculations give an excellent agreement with both RIXS and INS experiments. Note that Figure~\ref{fig:sw_contr} shows two interlaced branches around 25 and 90\meV. This is due to the fact that the unit cell contains two Ir and two Ni atoms. Only one branch actually carries significant physical intensity as shown in Figure~\ref{fig:sw}. \\

\begin{figure*}[H]
	\begin{center}
		\includegraphics[scale=0.5]{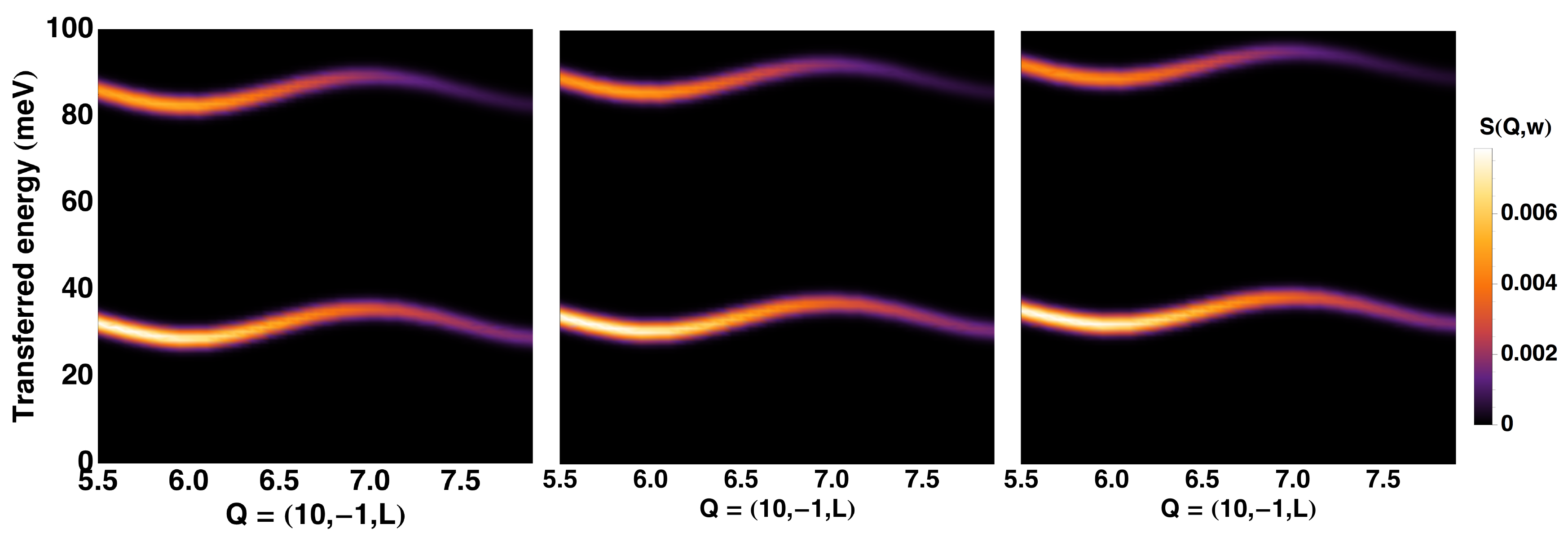}
		\caption{ \co Calculated spin waves with intrachain interactions $J_{xx}$ = 20\meV, $J_{zz}$ = 46\meV, magnetocrystalline anisotropy $D$ = 9\meV, and interchain interaction $J_{\perp}$ = -1\meV\ (left), 0\meV\ (middle) and +1\meV\ (right).} 
		\label{fig:Jinter}
	\end{center}
\end{figure*}

Using this program, we also examined the influence of the mean interchain interactions on the spin wave spectrum by calculating the response of three chains on a triangular lattice: two of them are antiferromagnetically coupled, while the third one is unfrozen in orientation. This spin configuration is one of the two that are compatible with the neutron diffraction experiment \cite{Lefrancois2014b}, the other one involving a modulation of the amplitude of the magnetic moments, which cannot be implemented in the {\it Spinwave} software. Note that  these two magnetic structures are equivalent up to a global phase. The inclusion of the interchain interactions modifies slightly the energy of the two branches (see Fig.~\ref{fig:Jinter}) by increasing (decreasing) their energy for ferromagnetic (antiferromagnetic) interchain coupling without affecting the amplitude of the dispersion. \\

The interchain exchange interaction might be estimated from the ordering temperature, but this is not straightforward for mixed-spin quantum chains where the intrachain exchange anisotropy competes with single-ion anisotropy so that the spin dimensionality evolves from almost XY to almost Ising by decreasing the temperature. An upper bound can nevertheless be perceived through analytical and numerical estimates from different models \cite{Oguchi1964,Scalapino1975,Villain1977,Boersma1981,Yasuda2005}. In the case of classical spin-S chains without anisotropy, by treating exactly the effects of the intrachain exchange $J_{\parallel}$ and in mean field those of the interchain exchange $J_{\perp}$, the ordering temperature is analytically computed in the form ${T_N \approx J_{\parallel}} S(S+1)((8/3) z_{\perp}J_{\perp}/J_{\parallel})^{1/2}$ where $z_{\perp}$ is the number of chains with which each chain interacts. With $S = 1/2$, $J_{\parallel} = J_{xx} = J_{yy} = 20$\meV\ and $T_N \approx 80$\K\ $\approx 6.9$\meV\ this leads to the exchange ratio $\eta=J_{\perp}/J_{\parallel} \approx 0.02$ and an interchain exchange $J_{\perp} \approx 0.4$\meV. With the larger exchange $J_{zz}$ a smaller exchange ration $\eta$ is obtained. N\'eel ordering is inhibited by interchain correlations and quantum fluctuations, which would suggest that $J_{\perp}$ is larger. Approached numerically these effects in the case of the extreme limit of S=1/2 quantum spins lead indeed to an increase of the exchange ratio up to $\eta \approx 0.1$ and therefore of interchain exchange up to $J_{\perp}\approx 2$\meV. These decrease to $\eta \approx 0.075$ and $J_{\perp}\approx 1.5$\meV\ for quantum spins S=3/2. N\'eel ordering on the other hand should be favored by the reduction of the spin dimensionality caused by the single ion and intra-chain exchange anisotropies that should lead to much lower values of the interchain exchange $J_{\perp}$, as suggested by the exchange ratio $\eta \approx 0.009$ obtained in the Ising limit.
It follows that the $J_{\perp} \approx 0.4$\meV\ estimated from the classical spin-S chains without anisotropy can be considered safely as providing an upper bound. \\

As a matter of fact, another independent estimate of $J_{\perp}$ can be obtained from the coercive field $\mathrm{H}_c$ which was observed up to $55$\T\ in the magnetization processes \cite{Singleton2014}. We assume that this coercive field corresponds to the minimum field of coherent reversal of spin chains. In case of the frustrated magnetic structure with two frozen chains of antiparallel moments over three with the third being paramagnetic, there is a degeneracy of the possible configurations. 
Each frozen chain can be surrounded by (3 + n) chains of opposite moment orientation and 6-(3+n) chains of same moment orientation, with n = 0, 1, 2, 3. On the other hand, each unfrozen chain is always surrounded by 3 chains of opposite moment orientation and 3 chains of same moment orientation. We can write $(3+\left<n\right>/3) J_{perp} \approx \mathrm{C~H}_c$ with C~$\approx 0.058$ for $\mathrm{H}_c$ given in Tesla and $J_{perp}$ in meV, which leads to values of $J_{perp}$ ($\approx$ 1.06\meV\, for $\left<n\right>= 0$, $\approx$ 0.80\meV\, for $\left<n\right> = 1$, $\approx$ 0.64\meV\, for $\left<n\right>= 2$ and $\approx$ 0.53\meV\, for $\left<n\right> = 3$) of the order of the upper bound extracted from the N\'eel temperature. Accordingly, the interchain exchange is expected to have a minor effect on the spin wave spectrum and confirms the quasi-one dimensional character of \SNIO.\\

We also calculated the effect of the intrachain Dzyaloshinskii-Moriya interactions, whose component along the {\it c} axis is allowed. We found that the small changes produced by this term in the energy position and the dispersion of the spin waves cannot be seen in the RIXS and INS experiments, up to values of 5 meV. Up to this upper bound, It has also no effect on the magnetic configuration.

\section{Monte Carlo calculations} 
Monte-Carlo simulations have been performed by considering a system of chains of alternating classical spin 1/2 and spin 1 sites. Interchain interactions have been neglected for simplicity. The exchange interactions and single ion anisotropies have been fixed to that provided by the spin-waves analysis. The simulations have been performed using a conventional Metropolis algorithm, cooling the system from 2000K down to 1K. The temperature was decreases exponentially (0.95 times the previous temperature). At each step in temperature, 10$^4$ spin flips per spin were conducted for equilibration and a further 10$^4$ steps for data taking. The magnetic susceptibility was calculated using 
the fluctuation-dissipation theorem:

\begin{equation}
\chi_\alpha=\frac{\langle m_{\alpha}^2 \rangle-\langle m_{\alpha} \rangle^2}{NT}
\end{equation}
where $N$ is the number of spins in the simulations and $\alpha$=x,y,z is the direction. Simulations using supercells of 3x3x200 up to 3x3x800 have been used. The results presented in the main article correspond to the largest supercell with 20000 spins.

\section{Magnetization measurements}

\begin{figure*}[H]
	\begin{center}
		\includegraphics[scale=0.75]{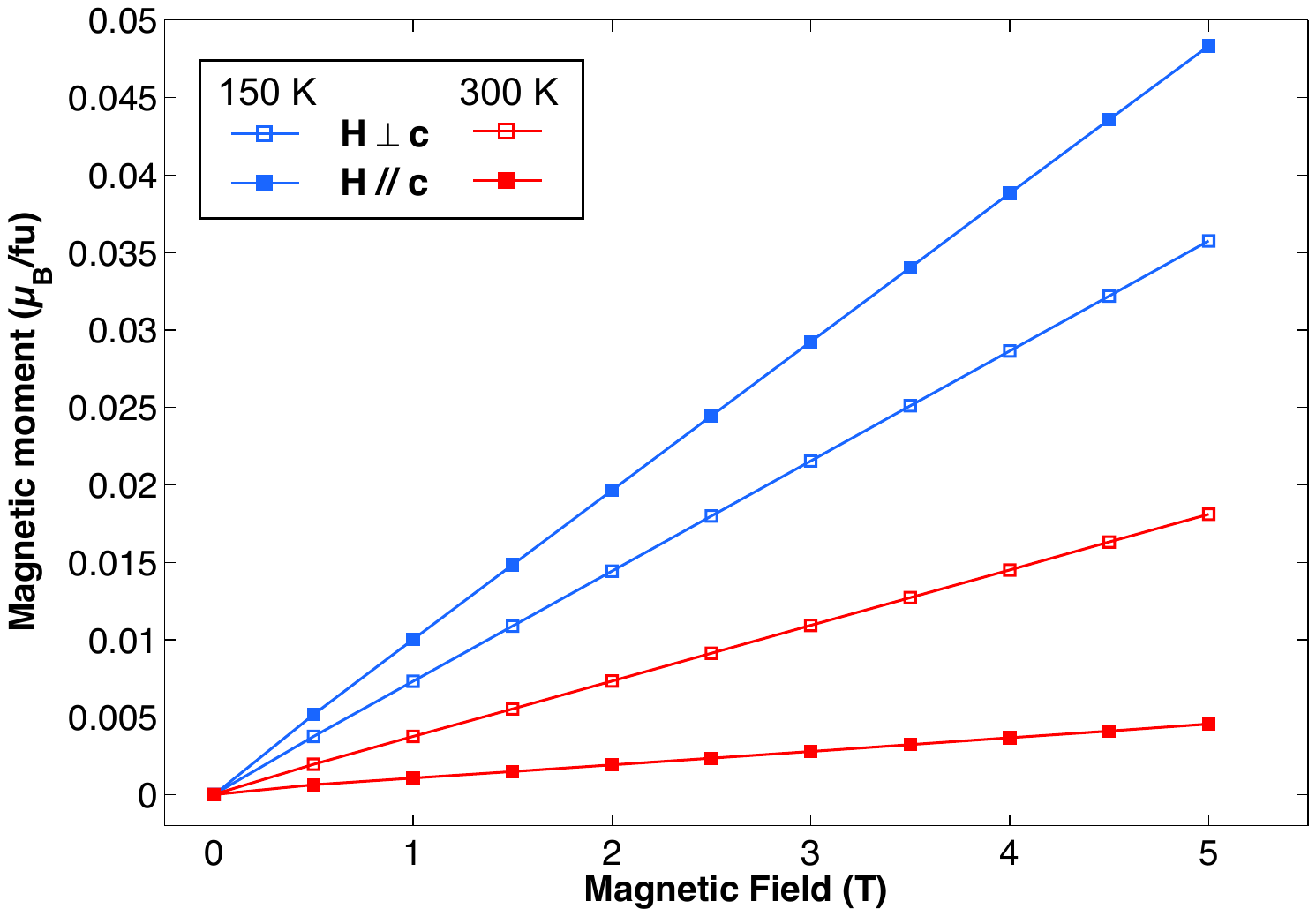}
		\caption{ \co Magnetic field dependence of the magnetization measured with the magnetic field applied parallel (filled symbol) and perpendicular (empty symbol)  to the $c$-axis at 300\K\ (in cyan) and 150\K\ (in red).} 
		\label{fig:MH}
	\end{center}
\end{figure*}

The easy-plane nature of the magnetocrystalline anisotropy of the compound can also be observed in the magnetic field dependence of the magnetization (see Fig.~\ref{fig:MH}). Indeed, at 300\K, the magnetization measured in the direction of the c axis is smaller than in the perpendicular direction. At 150\K, the inverse is observed.

\end{document}